\documentclass[final,3p,times,twocolumn]{elsarticle}
\biboptions{sort&compress}
\usepackage{graphicx,color}
\usepackage{latexsym,amssymb}
\newcommand{\bs}[1]{\rule[3.5mm]{0mm}{0mm}{\bfseries\textsf{#1}}}

\def\be{\begin{equation}}
\def\ee{\end{equation}}
\def\bc{\begin{center}}
\def\ec{\end{center}}

\def\no{N_{\rm open}}

\journal{Physica B}

\begin{document}
\begin{frontmatter}

\title{Disordered two-dimensional electron systems with chiral symmetry
\tnoteref{t1}}
\tnotetext[t1]{Dedicated to Costas Soukoulis on the occasion of his 60th birthday}
\author[rvt]{P. Marko\v{s}}
\ead{peter.markos@stuba.sk}
\author[focal]{L. Schweitzer}
\ead{ludwig.schweitzer@ptb.de}
\address[rvt]{%
Institute of Nuclear and Physical Engineering, FEI, Slovak University of
  Technology, 812\,19 Bratislava, Slovakia}
\address[focal]{Physikalisch-Technische Bundesanstalt (PTB), Bundesallee
  100, 38116 Braunschweig, Germany}

\begin{abstract}
We review the results of our recent numerical investigations on the
electronic properties of disordered two dimensional systems with
chiral unitary, chiral orthogonal, and chiral symplectic symmetry.  
Of particular interest is the behavior of the density of states and 
the logarithmic scaling of the smallest Lyapunov exponents in
the vicinity of the chiral quantum critical point in the band center
at $E=0$. The observed peaks or depressions in the density of states,
the distribution of the critical conductances, and the possible
non-universality of the critical exponents for certain chiral unitary
models are discussed.  
\end{abstract}

\begin{keyword}
Chiral symmetry, two-dimensional systems, electron localization 
\PACS 73.23.-b\sep  71.30.+h\sep  72.10.-d
\end{keyword}

\end{frontmatter}

\section{Introduction} 
Two-dimensional disordered systems have been attracting special
attention for many years because $d=2$ is the lower critical dimension 
of the metal-insulator transition (MIT) \cite{AALR79}. For lattice
systems with orthogonal symmetry (random on-site disorder with time
reversal symmetry) all electronic states are localized in the limit of
infinite system size. However, for weak disorder and energies close to
the band center, the localization length can become very large. On
length scales smaller than the localization length, the wavefunctions
exhibit self-similar (fractal) behavior \cite{SE84}.  
The presence of spin dependent hopping changes the symmetry of the
model to symplectic, and enables the system to undergo a
metal-insulator transition at a certain value of the disorder 
strength \cite{HLN80,EZ87,And89,ASO02,ASO04}. The critical eigenstates
at the MIT exhibit multifractal properties \cite{Sch95} and the   
localization length was reported to show a parity dependence
\cite{ASO03}. 
A strong magnetic field turns the symmetry to unitary and induces
critical states \cite{SKM84a}, i.e., singular energies where the
localization length of the multifractal eigenstates \cite{HKS92,HS94}
diverges, which are important for the explanation of dissipative
transport \cite{WLS98,SM05} in the quantum Hall effect.   

Two-dimensional (2D) models possessing an additional chiral symmetry
exhibit various electronic properties not observed in the situations
mentioned above. The chiral symmetry can be found in models defined
on bi-partite lattices with non-diagonal disorder only
\cite{ITA94,EK03}. Despite the disorder, the energy eigenvalues appear
in pairs, $E_n$ and $-E_n$ symmetrically around the band center $E=0$
(for the definition of chiral 2D models, see section \ref{models}). 
Chiral symmetry implies unusual properties of the model in the
vicinity of the band center $E=0$. For most chiral cases, the density
of states and the localization length are diverging and the band
center is a quantum critical point \cite{GW91,Gad93}. 
At zero temperatures, an infinite sample is metallic at $E=0$ but
insulating for any non-zero energy. The appearance of the criticality
at $E=0$ originates from the chiral symmetry. However, the existence
of the critical point also depends on the boundary conditions. As
listed in Table~\ref{symtab}, the sample exhibits chiral symmetry only
for special combinations of boundary conditions and parity. This
boundary and parity dependence of the sample's length $L_z$ and width
$L$ has no analogy in `standard' disordered models. 

\begin{table}[t]
\bc
\begin{tabular}{l|c|c|c||c|c|c|}
$L_z\ \setminus\ L$ & \multicolumn{3}{c}{odd}\vline\hspace*{1.7pt}& 
\multicolumn{3}{|c|}{even} \\  
\cline{1-7}
& & $D_x$&$P_x$ &  &   $D_x$&$P_x$   \\ 
\cline{2-7} 
odd & $D_z$ & \bs{Ch+} & \bs{} &   $D_z$ &\bs{Ch} & \bs{Ch}   \\ 
\cline{2-7} 
         & $P_z$ &  \bs{}  &  \bs{}   & $P_z$ & \bs{}  &  \bs{} \\ 
\cline{1-7}\\[-10.5pt]
\cline{1-7}
 & & $D_x$  & $P_x$ &   &  $D_x$&$P_x$   \\ 
\cline{2-7}
even     & $D_z$ & \bs{Ch} &  \bs{} &  $D_z$& \bs{Ch} & \bs{Ch} \\ 
\cline{2-7} 
         & $P_z$ & \bs{Ch}  &  \bs{}  & $P_z$ & \bs{Ch} & \bs{Ch} \\ 
\cline{1-7}
\end{tabular}
\ec
\caption{Two dimensional models with non-diagonal disorder possess
  the chiral symmetry only for special choices of the boundary
  conditions and the parity in the number of sites. For a given
  combination of boundary conditions, periodic (P) and Dirichlet (D),
  the chiral symmetry (\textsf{Ch}) and chiral symmetry with an extra
  eigenvalue at $E=0$ (\textsf{Ch+}) is observed
  \cite{MW96,BMSA98,AS99,MBF99}.} 
\label{symtab}
\end{table}

The special symmetry of the energy spectra may be accompanied by a
non-analytical behavior of the density of states (DOS) at the chiral
critical point \cite{Gad93,FC00,MDH02,EM08}. 
In 2D chiral unitary models defined on a bricklayer
\cite{SM08a}, which has the same topology as graphene's honeycomb
lattice, the DOS exhibits a sharp drop near the band center going to
zero at $E=0$ and depends on both disorder and system size
\cite{Sch09}. Contrary to this behavior, the DOS of the 
chiral orthogonal system is finite at the band center and showing a
narrow extra peak in the case of square lattice samples (see below).   

Similarly to other critical regimes, systems with chiral symmetry  
can be analyzed using the single parameter scaling theory
\cite{AALR79}. However, the scaling parameter is not the ratio of the
system size $L$ to the correlation length $\xi(E)$, but the ratio of
the logarithm of these parameters instead \cite{SH02}
\be\label{log-scal}
\chi = \frac{\ln L}{\ln \xi(E)/\xi_0}.
\ee
Also, the energy dependence of the correlation length is logarithmic
\cite{GW91,Gad93,FC00},  
\be\label{log-kappa}
\ln(\xi(E)/\xi_0) \sim |\ln (E_0/|E|)|^\kappa,
\ee
in contrast to the power-law scaling dependence $\xi(E)\sim
|E-E_c|^{-\nu}$ observed in non-chiral disordered systems. Thus,
models with chiral symmetry enable the detailed analysis of
logarithmic scaling, discussed previously in~\cite{SH02}.

\begin{figure}[t!]
\bc
\includegraphics[clip,width=0.4\textwidth]{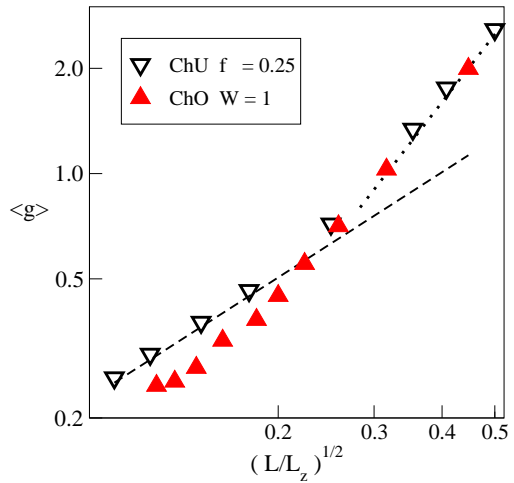}
\ec
\vspace*{-3mm}
\caption{The length dependence of the mean conductance $\langle
  g\rangle$ for quasi-one dimensional chiral systems of length
  $L_z$ and fixed width $L=65\,a$. The energy is $E=0$. Due to the
  chiral symmetry of the model, the crossover from the Ohmic
  $1/L_z$-behavior (dotted line) to the $1/\sqrt{L_z}$ dependence
  (dashed line) is observed, in agreement with theoretical predictions
  \cite{MBF99}. Two models with chiral unitary (ChU: $f=0.25\,h/e$)
  and   chiral orthogonal (ChO: $W/t_0=1$) symmetry, defined in
  section \ref{models}, were considered.   
}
\label{fig-cond}
\end{figure}

Chirality also strongly affects the transport properties of the system. 
The non-Ohmic behavior of chiral systems with an odd number of open
channels, where the mean conductance decreases much more slowly with
the length of the system (see Fig.~\ref{fig-cond}), has been predicted
theoretically \cite{MBF99} and confirmed numerically in~\cite{SM08a,MS07}.  

In this paper we review some results of our recent investigations on
electric transport properties of two dimensional chiral systems.  
In the following section we briefly introduce the models studied. 
In section~\ref{results}, we summarize our findings for the Lyapunov
exponents, the critical conductance and its probability distribution,
and show recent calculations for the density of states. Of special
interest is the scaling analysis of the diverging critical electronic
states at $E=0$ \cite{MDH02,MRF03,HWK97}.

\section{Models}\label{models}
In the absence of diagonal disorder the single-band tight-binding
Hamiltonian defined on the sites {$n$} of a two-dimensional bricklayer 
or square lattice with lattice constant $a$ reads 
\be
{\cal H}=\sum_{\langle n\ne n'\rangle} t_{nn'} c_n^{\dagger}c_{n'}^{},
\ee
where the sum is over nearest neighbors only. The random disorder is
incorporated in the hopping terms, which also determine the symmetry
of the problem. Square lattice and bricklayer lattice differ only in
the absence of every other vertical bond in the latter and so the
coordination number is reduced to three (see Fig.~\ref{fig-lattice}).

\subsection{Unitary symmetry}
To describe a disordered chiral 2D system with broken time-reversal
symmetry, the hopping terms in the (transversal) $x$
direction are chosen to acquire complex phases and are defined as 
\be\label{eq-u}
t_x = t_0 e^{i\theta_{x,z;x\pm a,z}}, 
\ee
where for a bricklayer lattice the phases
$\theta_{x,z;x+a,z}=\theta_{x,z+2a;x+a,z+2a}-\frac{2\pi e}{h}\Phi_{x,z}$ 
are determined by the total flux threading the plaquette at $(x,z)$
\be
\Phi_{x,z} = \frac{p}{q}\frac{h}{e} +\phi_{x,z}.
\ee
Here, $p$ and $q$ are mutual prime integers and the magnetic flux
density perpendicular to the two-dimensional lattice $B=ph/(qe2a^2)$
is described by the number $p/q$ of magnetic flux quanta $h/e$ per
plaquette $2a^2$. This differs from the random flux model studied
previously \cite{MS10}, where the constant magnetic field part was
absent. The random part is generated by the local fluxes $\phi_{x,z}$,
which are uniformly distributed $-f/2\le\phi_{x,z}\le f/2$ with zero
mean. The disorder strength $f$ can be varied within the interval from
$f/(h/e)=0$ to $f/(h/e)=1$.   

\subsection{Orthogonal symmetry}
For the chiral orthogonal symmetry, we consider two different models.
Both are defined on a square lattice. In the first, we set $t_z/t_0=1$ 
and take the disordered hopping terms as random real numbers that are
defined as 
\be\label{eq-ty}
t_x = t_0\exp \frac{W}{t_0}\varepsilon,
\ee
where \{$\varepsilon$\} is a set of uncorrelated random numbers with
box probability distribution $|\varepsilon|\le 1/2$. The strength of
the disorder was varied in the range $W/t_0=2$ to $W/t_0=10$. 
In the second model, the transfer terms in both the $x$ and $z$
direction are random numbers which are box-distributed about $t_0$
within the interval [$t_0-W_s/2$, $t_0+W_s/2$] with possible disorder
strengths $0\le W_s/t_0\le 2$. In the latter model, single sites may get
isolated and decoupled from the remaining 2D lattice for $W/t_0 \to 2$.

\subsection{Symplectic symmetry}
The off-diagonal disorder as given by Eq. (6) with $W/t_0=4$ was used
in the chiral symplectic model, where we consider a chiral version of
the Ando model \cite{And89,MS06}. The hopping terms $t_{nn'}$ are now
$2\times 2$ matrices   
\be
t_\parallel = t_0\left(
\begin{array}{ll}
c & -s \\
-s &   c
\end{array}
\right), \quad
t_\perp = t_x\left(
\begin{array}{ll}
c & is \\
-is &   c
\end{array}
\right),
\ee
where $c^2+s^2=1$, $s=1/2$, and the disorder in the hopping $t_x$
is given by Eq.~(\ref{eq-ty}).

\subsection{Lattice topology}
\begin{figure}
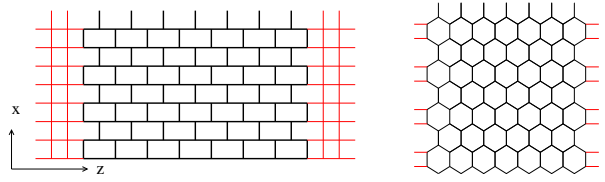

\includegraphics[clip,width=0.28\textwidth]{ms_pb_fig2a.eps}\hfill
\includegraphics[clip,width=0.15\textwidth]{ms_pb_fig2b.eps}
\caption{The bricklayer lattice (left) shares the topology of the
  honeycomb lattice (right). The red lines display the attached
  perfect semi-infinite leads used in the calculation of the scaling
  variables $z_i$ and the two-terminal conductance.} 
\label{fig-lattice}
\end{figure}

The two-dimensional lattices of size $L\times L_z$ on which the above 
models are defined, are either a regular square lattice or a
bricklayer lattice which mimics the honeycomb lattice 
(Fig.~\ref{fig-lattice}). The width $L$ and the length $L_z$ of the
lattice are taken to be similar $L_z\simeq L$ for the investigations 
of the DOS and for the conductance, whereas quasi-one-dimensional
samples with $L_z\sim 10^9$ lattice units are used in the analysis
of the scaling properties of the models.

\section{Numerical methods}
For the numerical calculation of the two-terminal conductance, we
attach two ideal (without disorder) semi-infinite leads to the 
finite sample having width $L$ and length $L_z$ and apply the well
known Economou-Soukoulis formula \cite{ES81}  
\be
g = \frac{e^2}{h}\, \textrm{Tr}~T^\dagger T = \frac{e^2}{h}\sum_{i=1}^{\no}
\frac{1}{\cosh^2 (x_i/2)}, 
\ee
where $T$ is the corresponding transmission matrix and the $x_i$
parameterize the eigenvalues of the hermitian matrix $T^\dagger T$. The
parameter $\no$ determines the number of open channels in the attached
leads. Owing to the disorder, the conductance is a statistical
variable. Therefore, an ensemble of finite samples, which differ only
in the microscopic realization of the disorder, is considered and the  
mean value, the variance, and probability distribution of the
conductance is evaluated using the algorithm described in
\cite{PMR92}.   

We define the dimensionless quantities $z_i=x_i L/L_z$ to be used for
the scaling analysis in the vicinity of the critical point \cite{KM93}. 
In the limit $L_z\gg L$, the ratio $z_1/(2L)$ converges to the smallest 
Lyapunov exponent $\gamma=\lim_{L_z\to\infty} x_1/(2L_z)$ where
$x_1$ is the smallest positive eigenvalue of $\ln(T^\dagger T)$. 
We calculated numerically the first two parameters $z_1$ and $z_2$ for 
Q1D systems of length up to $L_z\sim 10^8 L$  in order to achieve a
relative uncertainty for $z_1$ of order of $10^{-4}$. 
The data obtained for $z_1(E,L)$ and $z_2(E,L)$ were fitted to
scaling formulae (\ref{log-scal}) and (\ref{log-kappa}).

\section{Properties of chiral systems}\label{results}
In this section, we summarize our numerical results obtained for
various disordered systems obeying chiral symmetry.

\subsection{Scaling variables $z_i$}
The calculated spectrum of scaling variables $z_i=x_i L/L_z$ is
strongly influenced by the presence of the chiral symmetry
\cite{BMSA98,MBF99}. They take on also negative values for chiral
unitary systems. Our data \cite{MS07} confirmed that    
\be
|z_{2a-1}| = |z_{2a}| = c\left[a-1/2\right]
\ee
for an even number of open channels, and
\be\label{LE-DBC}
|z_i| = \left\{
\begin{array}{ll}
c\ {\rm Int}[a/2]    &  {\rm Dirichlet~BC }\\
(c/2)\ [a-1/2]    &  {\rm periodic~BC}
\end{array}\right.
\ee
for an odd number of channels. From Eq. (\ref{LE-DBC}) it follows that
the mean value of the smallest $z_1$ is zero.  

\begin{figure}
\bc
\includegraphics[clip,width=0.45\textwidth]{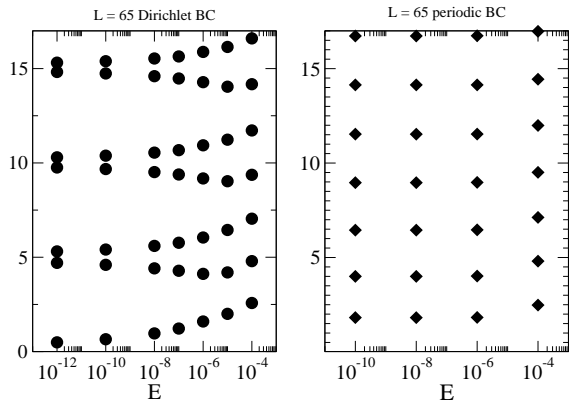}
\ec
\caption{The spectrum of scaling variables $|z_i|$ for the Q1D
  systems with chiral orthogonal symmetry. A lattice $L\times
  L_z$ with square topology is considered with an odd value $L=65\,a$
  and two boundary conditions (BC) Dirichlet (left) and periodic
  (right) imposed in the transversal direction. For Dirichlet BC, the
  model is chiral orthogonal at the band center $E=0$ and
  $z_1(E=0)=0$, in agreement with Eq.~(\ref{LE-DBC}). The symmetry
  changes to orthogonal for non-zero energy $E$. Note the degeneracy
  in the spectrum of scaling variables for the Dirichlet BC. This
  degeneracy is   broken for non-zero energies $E$. There is no chiral
  symmetry for periodic BC. The disorder strength in the vertical
  hopping (\ref{eq-u}) is $W=5$. Similar results for the chiral
  unitary symmetry were published in \cite{MS07}. 
}
\label{figr1}
\end{figure}

The same relations hold also for chiral orthogonal systems. 
Figure \ref{figr1} shows the spectrum of scaling variables for 
$\no$ odd and two types of boundary conditions. The numerical data
indeed confirm that $z_1=0$ when Dirichlet BC are imposed. This is a
reason for the non-Ohmic behavior of the mean conductance shown in
Fig.~\ref{fig-cond}. 
Also,  Fig.~\ref{figr1} shows that the degeneracy of the spectrum of
absolute values $|z_i|$ is lifted for any non-zero energy. A more
detailed analysis of the scaling variables $z_i$ for the chiral
unitary system is given in Ref.~\cite{MS07}.  

\begin{figure}
\bc
\includegraphics[clip,width=0.4\textwidth]{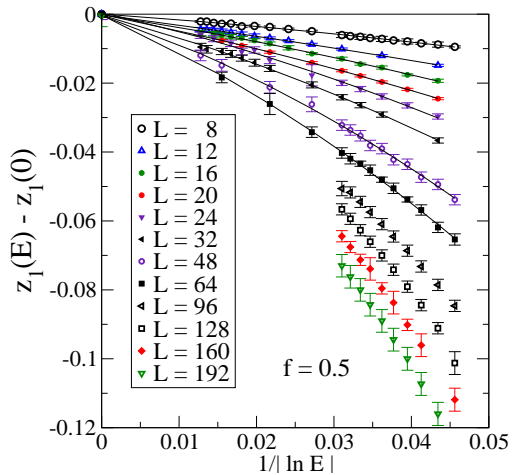}
\ec
\caption{The smallest scaling variable $z_1$ for 
  chiral unitary systems without a constant magnetic field, calculated
  on the bricklayer lattice. The number of open channels $\no$ is
  identical with the width of the layer. Periodic BC were used in the
  transversal direction. Note that the energy decreases up to $E/t_0\sim
  10^{-30}$. Since quartic precision in the numerical calculations is
  necessary to treat such energies, we were able to calculate the
  scaling variables $z_i$ for these energies only for narrow samples
  ($L\le 24$), due to the accuracy and CPU requirements. 
}
\label{unit-z1z2}
\end{figure}

\subsection{Scaling analysis}
Scaling theory predicts that in the vicinity of the critical point
$E=0$, the smallest scaling variable $z_1$ is a function of only one
parameter, 
\be
z_1(E,L) = F(\chi)
\ee
where $\chi = \ln L/\ln(\xi(E)/\xi_0)$, and $\xi(E)$ is the
correlation length 
\be\label{xiE}
\ln \frac{\xi(E)}{\xi_0} \sim \left|\,\ln
\left(\frac{E_0}{|E|}\right)\,\right|^\kappa
\ee
with unknown energy parameter $E_0$ and exponent $\kappa$.

Figure \ref{unit-z1z2} shows typical numerical data obtained
in the vicinity of the chiral critical point. The data
confirm that both $z_1$ and $z_2$ can be described as a function 
of $\ln E$. However, a logarithmic energy dependence of $\xi(E)$,
given by Eq.~(\ref{xiE}) can be observed only for very small values of
the energy, typically $|E/t_0|<10^{-8}$.   

\begin{figure}
\bc
\includegraphics[clip,width=0.42\textwidth]{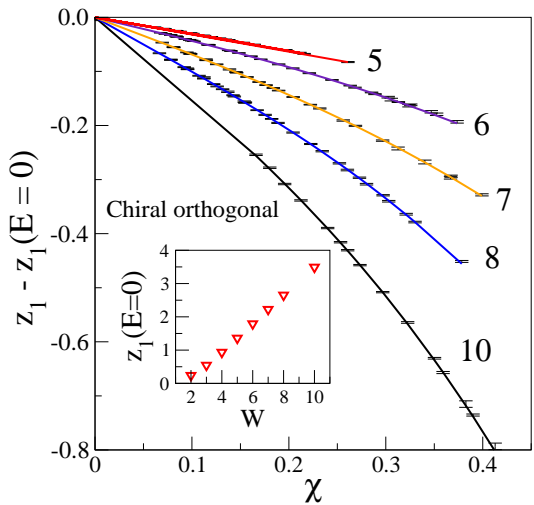}
\ec
\caption{Scaling of the smallest scaling variable $z_1$ for 
  chiral orthogonal models with hopping terms $t_x$ given by
  Eq.~(\ref{eq-ty}). The disorder strength $W/t_0$ is given in the
  Figure. The scaling variable $\chi$ on the horizontal axis is given
  by Eqs.~(\ref{log-scal}) and (\ref{log-kappa}) with $\xi_0=1$,
  $E_0\approx -4.5\,t_0$ and the exponent $\kappa$ close to the value
  2/3 \cite{SM11}. The inset shows the disorder dependence of
  $z_1(E=0)$ for $L=96\,a$. 
}
\label{beta-1-W}
\end{figure}

For orthogonal and symplectic models as described in section
\ref{models} with disorder given by (\ref{eq-ty}), we verified the
scaling behavior (\ref{xiE}) for each given disorder strength. 
In all systems we used periodic BC and even $\no$. This choice enables 
us also to compare systems with and without a constant magnetic
field. Systems with odd $\no$ and Dirichlet BC are numerically not
accessible since the smallest $z_i$ is zero for $E=0$ and very small
for non-zero energy (Fig.~\ref{figr1}).  

We fit the numerical data for $z_1$ and $z_2$ to the polynomial
function of $\chi$ and extract the critical exponent $\kappa$. 
We found that for all models $\kappa\approx 2/3$, in agreement with
previous theoretical predictions \cite{HWK97,MDH02,MRF03}. As an
example, we show in Fig.~\ref{beta-1-W} the result of our scaling
analysis for the chiral orthogonal model and various strengths of the
disorder. The disorder dependence of $z_1(E=0,W)$ is plotted in the
inset. This shows that the the value of the disorder $W$ represents 
an extra parameter that defines the respective model.  

Although the addition of a constant magnetic field to the chiral
orthogonal model breaks the time-reversal symmetry, it does not
influence the value of the exponent $\kappa\approx2/3$.  The only
exception is the chiral unitary model without constant magnetic field
where $\kappa\simeq 1/2$ \cite{MS10} (the raw data for this model are
shown in Fig.~\ref{unit-z1z2}). More details about the scaling at
chiral quantum critical points can be found elsewhere \cite{SM11}.
 
We did not succeed in obtaining the logarithmic scaling from the raw
data for the mean conductance \cite{MS07,SM08a}. The reason is that for
each particular sample the conductance is given as a sum of
contributions of all channels. Since the sum $z_1+z_2$ is almost
constant also for non-zero energies, the conductance depends only very
weakly on the energy in the vicinity of the critical point. This tiny  
energy dependence, if any at all, cannot be extracted from our
numerical data.

\subsection{Conductance}
\begin{figure}
\bc
\includegraphics[clip,width=0.42\textwidth]{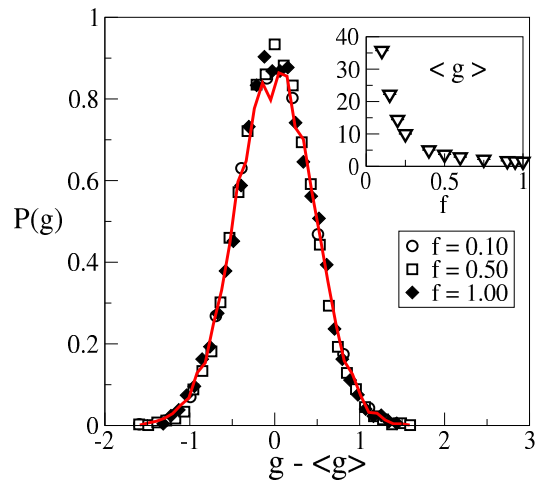}
\ec
\caption{The probability distribution $P(g-\langle g\rangle)$ for the
  chiral unitary system without magnetic field and three different
  values of $f$. The width of the distribution var\,$g \approx 0.187$
  does not depend on $f$. The solid line is the conductance distribution
  for the chiral unitary model with $f=0.5\,h/e$ and with an additional
  magnetic field $B=(1/12)\,h/(ea^2)$ ($L/a=72$, mean conductance
  $\langle g\rangle = 2.82$, var\,$g=0.207$). The inset shows the mean
  conductance $\langle g\rangle_c$ for various values of the random
  field $f$. Square samples of size $257\times 257$ were considered
  with Dirichlet BC in the transversal direction. Note that the
  strength of the disorder is limited in the present model. Even for
  $f=1\,h/e$, we get a rather large mean conductance, $\langle
  g\rangle = 1.45\,e^2/h$. 
}
\label{figr11}
\end{figure}

At the critical point, the 
probability  distribution of the conductance 
$P(g)$ does not depend on the size of the system, and
possesses the typical shape for a given universality class. 
In Fig.~\ref{figr11} we show the conductance distribution for the
chiral unitary system without a constant magnetic field. The data
confirm that the distribution does not depend on the strength of the
disorder $f$, which in this case measures the fluctuation of the flux
through the plaquettes. Since the disorder is weak, the mean value of
the conductance is large, and decreases when the disorder increases,
as shown in the inset. The critical distribution is Gaussian with a
universal width, var\,$g=0.187$.  

Of particular interest is the comparison of the two unitary chiral
systems that differ from each other only by the presence of a constant
magnetic field. As discussed in the previous section, these two
systems exhibit a different critical exponent $\kappa$. Surprisingly,
the conductance distributions for these two models are almost
identical as shown in Fig.~\ref{figr11}. Also, the mean conductance is
larger than 1 even in the case of strongest disorder possible
($f=1\,h/e$). Therefore, our chiral unitary model does not allow
studies of strongly disordered chiral systems. 

\begin{figure}
\bc
\includegraphics[clip,width=0.4\textwidth]{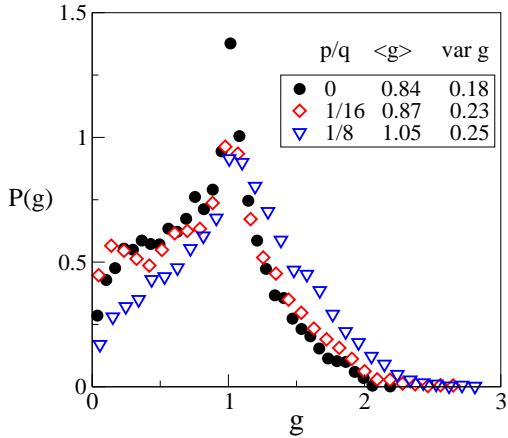}
\ec
\caption{The conductance distribution for the two-dimensional
  disordered chiral orthogonal model on a square lattice with random
  hopping terms (\ref{eq-ty}), and with a constant magnetic field,
  $B=(p/q)\,h/ea^2$.  The disorder strength is $W/t_0=5$, and the
  size of the system $L/a=64$. The legend gives the ratio $p/q$ which
  determines the magnetic field and corresponding mean conductance and
  the variance. Note the typical non-analytical form of the distribution
  at $g=1\,e^2/h$   
\cite{Mar06}.
}
\label{brick-ou}
\end{figure}

Smaller values of the critical
conductance can be obtained for systems with random hopping terms as
given by Eq.~(\ref{eq-ty}). Figure~\ref{brick-ou} shows the critical
conductance distribution for system with real hopping terms given by
Eq.~(\ref{eq-ty}) with disorder $W=5$ and various strengths of the 
constant magnetic field. While the system without magnetic field
belongs to the chiral orthogonal universality class, the constant
magnetic field changes the symmetry to chiral unitary. In spite of the
different symmetry classes, no significant difference between the
critical distributions is observed. Also the critical exponent
$\kappa\approx 2/3$ is the same for both models \cite{SM11}.

\subsection{Density of states}
In many `standard' disordered models, the density of states can be
considered to be constant in the critical region. A very strong
energy dependence of the DOS can affect the accuracy of the numerical 
scaling. A typical example is the metal-insulator transition in the
band tails (see, for instance, \cite{BM06}). Recently it was found
that chiral unitary models can exhibit a very narrow dip in the DOS at
the band center \cite{Sch09}. It has been suggested \cite{AS99,SA01}
that these `microgaps' appearing in disordered chiral models are the
consequence of the non-perturbative ergodic regime. This can happen
for times large compared with the diffusion time when the localization
length exceeds the sample size. 

\begin{figure}
\includegraphics[clip,width=0.47\textwidth]{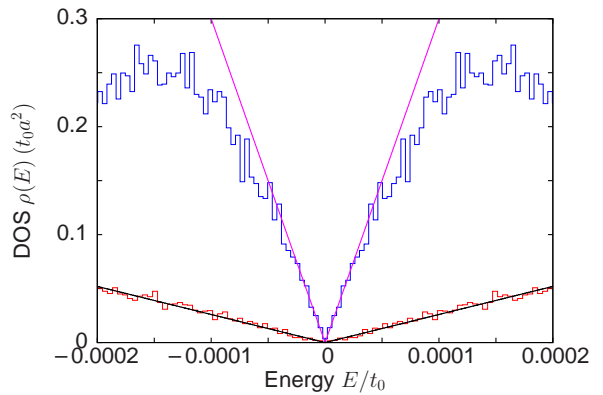}
\caption{The depression in the density of states $\rho(E)$ near the
  quantum critical point at $E=0$ for the random magnetic flux model
  with (upper blue curve) and without (lower red curve) a spatial
  constant magnetic flux density defined on a bricklayer lattice. 
  The histogram bin-width is about $4\times 10^{-6}$. The dip in the
  DOS depends on both the disorder strength and the sample 
  size. The brown and magenta solid lines are fitting curves
  \cite{Sch09}. The RMF strength and system size are $f=0.5\,h/e$,
  $L=L_x=96\,a$ for $B=0$, and $f=0.5\,h/e$, $L=L_x=120\,a$ in the
  case of $B=(1/12)\,h/(2ea^2)$, respectively.}  
\label{unidos}
\end{figure}

As an example, the DOS in the vicinity
to the chiral unitary critical point is shown in
Fig.~\ref{unidos}. The energy range of the DOS depression gets
narrower with increasing RMF disorder strength for zero constant
magnetic field but it becomes broader in the case of a finite constant
magnetic field \cite{Sch09}. In the latter case, two additional
quantum Hall critical points (QHCPs) showing the usual power-law
divergence with critical exponent $\nu\simeq 2.4$ \cite{SM08a} exist
symmetrically about the chiral point at $E=0$. Within the achieved
uncertainty this value for the chiral quantum Hall case obtained for a
bricklayer lattice is compatible with the recent high-precision
estimates $\nu\simeq 2.6$ \cite{SO09,AMSD11} obtained for the
Chalker-Coddington network model \cite{CC88}.    
With increasing disorder, the QHCPs move further apart
\cite{SM08a} and so they can be easily distinguished from the chiral
one remaining at $E=0$ and studied here. 

\begin{figure}
\bc
\includegraphics[clip,width=0.37\textwidth]{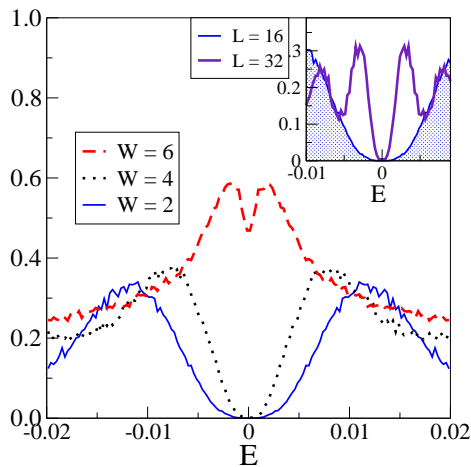}
\ec
\caption{The density of states for the chiral symplectic model
  calculated by diagonalizing $2\times 10^5$ samples of size
  $16\times 16$. A dip in the density of states is present only for 
  weak disorder and small size of the lattice,  but disappears when
  disorder and/or lattice size  increases. The inset shows the density
  of states for disorder $W =2$ and $L=16$ and $L=32$. 
}
\label{ando-dos}
\end{figure}

\begin{figure}
\bc
\includegraphics[clip,width=0.4\textwidth]{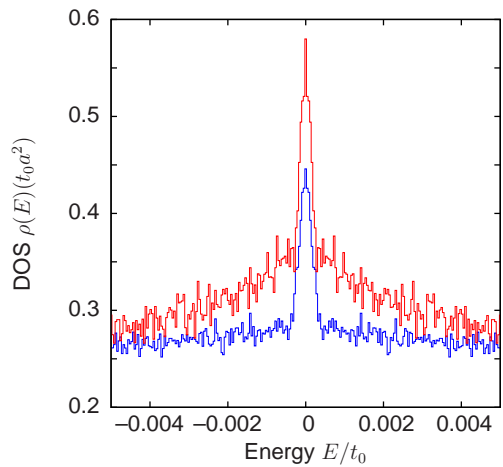}
\ec
\caption{The density of states for the chiral orthogonal model
  defined on a square lattice with real random hopping in both
  directions obtained by diagonalization of more than $2\times 10^4$
  samples of size $80\times 80$ and disorder strengths $W_s/t_0=1.0$
  (blue) and $W_s/t_0=1.5$ (red). No dip but an extra peak is observed
  in the  density of states close to the quantum critical point $E=0$.  
}
\label{ortho-dos}
\end{figure}

To check whether this is a general feature of systems with chiral
symmetry, we calculated numerically also the density of states for
chiral orthogonal and for chiral symplectic models. The results are
shown in Figs.~\ref{ando-dos} and \ref{ortho-dos}. 
Similar to the case of chiral unitary symmetry, the chiral symplectic 
system shows a spectral gap in the DOS at $E=0$. For a square system
$L=64$ and $W=3$ hopping disorder in both spatial directions, the
narrow gap is still visible. With increasing disorder strength or
system size, however, the gap becomes narrower until it seems to
disappear due to the insufficient numerical resolution.   

In the chiral orthogonal  case, instead of a dip, an extra peak is
seen on top of the 
disorder broadened DOS at the quantum critical point $E=0$. This
result was obtained for a square lattice model with random hopping in
both directions where van Hove singularities appear at the band center
and not at $E/t_0=\pm 1$ as for the bricklayer (hexagonal) lattice. An
explanation for the occurrence of the extra peak is the possible
isolation of single lattice sites that become disconnected from the
remaining system with increasing disorder. These sites contribute to
the DOS with eigenvalues close to zero.

\section{Conclusion}
Two dimensional systems with chiral symmetry exhibit interesting new
physical properties, not observed in standard disordered models. The
most striking feature is that the existence of a chiral quantum
critical point is determined by the boundary conditions and by the
parity of the size of the system. The chiral critical point coincides
with narrow structures in the energy dependence of the density of
states which depend both on the disorder strength and on the system
size. To study the intriguing scaling behavior, a logarithmic scaling
ansatz is necessary which replaces the usual power-law dependence
applied in ordinary Anderson localization with diagonal disorder.    

Owing to the logarithmic energy dependence of the correlation length,
the critical region is very narrow. Typically, it is narrower than the
$10^{-8}$th part of the bandwidth. This makes the scaling analysis of
the chiral critical behavior rather difficult. To achieve relevant
data with sufficient accuracy, extremely long systems must be
calculated, in some cases with quartic precision (up to 34 digits) of
arithmetic operations. This is probably the reason why the logarithmic
scaling was not observed in previous numerical studies
\cite{MS07,Cer00,Cer01,ERS01,ER04}. Our numerical data indicate that
all chiral systems can be divided into two classes: the first one
consists of only one model with random magnetic flux and zero magnetic
field. For this model, we observed that the critical exponent which
governs logarithmic energy dependence of the correlation length is
$\kappa\simeq 1/2$. For all other models, the critical exponent is
$\kappa\simeq 2/3$, independent of the symmetry. 

\bigskip
\noindent{PM thanks  Project VEGA 0633/09  for financial support.}


\end{document}